\input{aipcheck}

\documentclass[
    ,final            
  ]
  {aipproc}

\layoutstyle{6x9}

\begin{document}

\title{Electronic properties of graphite in tilted magnetic fields}


\author{Nataliya A.\ Goncharuk}
{
}
\author{Ludv\'{i}k Smr\v{c}ka}{
  address={Institute of Physics, Academy of Science of the Czech
Republic, v. v. i.,\\ Cukrovarnick\'{a} 10, 162 53 Praha  6, Czech Republic}
}

\begin{abstract}
The minimal nearest-neighbor tight-binding model with the Peierls
substitution is employed to describe the electronic structure of
Bernal-stacked graphite subject to tilted magnetic fields. We show
that while the presence of the in-plane component of the magnetic
field has a negligible effect on the Landau level structure at the K
point of the graphite Brillouin zone, at the H point it leads to the
experimentally observable splitting of Landau levels which grows
approximately linearly with the in-plane field intensity.
\end{abstract}
\maketitle

\section{Introduction}
Recently, graphene monolayers have attracted much attention motivated by
their unusual two\discretionary{-}{-}{-}dimensional (2D) Dirac energy spectrum of electrons. In graphite, which is a graphene multilayer composed of
weakly coupled graphene sheets, the interlayer interaction converts
the 2D electron energy spectrum of graphen into the three-dimensional
(3D) spectrum of graphite. The application of the tilted magnetic
field is a typical method to distinguish between 2D and 3D electron
systems, as in 3D systems the orbital effect of the in-plane magnetic
field should be observable. This problem has been touched in two
recent theoretical studies of the graphene multilayer energy spectrum
in magnetic fields parallel to the layers \cite{Pershoguba_2010}, and
of the Landau levels (LLs) of bilayer graphene in magnetic fields of
arbitrary orientation \cite{Hyun}. Both papers conclude that the very
strong in-plane field components are necessary to induce observable
effects on the electronic structure. Here we show that at the H point
of the hexagonal Brillouin zone of graphite the application of the
tilted magnetic field leads to experimentally observable splitting of
LLs.
\section{Model and results}
Bulk graphite is composed of periodically repeated graphene bilayers
formed by two nonequivalent Bernal-stacked graphene sheets. There are
two sublattices, A and B, on each sheet and, therefore, four atoms in
a unit cell. The distance between the nearest atoms A and B in a
single layer is $1.42\,$ {\AA}, the interlayer distance between
nearest atoms A is $d=3.35\,${\AA}.

To describe the graphite band structure, we employ the minimal
nearest-neighbor tight-binding model introduced in
Ref.~\cite{Koshino_2008} and successfully applied in
Refs.~\cite{Pershoguba_2010,Hyun,Orlita_2009}. The wave functions are
expressed via four orthogonal components $\psi^A_j$, $\psi^B_j$,
$\psi^A_{j+1}$, $\psi^B_{j+1}$, which are, in zero magnetic field,
Bloch sums of atomic wave functions over the lattice sites of
sublattices A and B in individual layers. The tight-binding
Hamiltonian $\mathcal{H}$ includes only the intralayer interaction
$\gamma_0=3.16\, \rm{eV}$ between nearest atoms A and B in the
plane, and the interlayer interaction $t=0.38\, \rm{eV}$ between
nearest atoms A out of plane.
The continuum approximation is used in the vicinity of the $H - K - H$
axis of the graphite Brillouin zone, for small $\vec{k}$ measured from
the axis. Then the electron wavelength is larger than the distance
between atoms, and the non-zero matrix elements of $\mathcal{H}$ can
be written as
\begin{equation}
\mathcal{H}^{AB}=\hbar v_F(k_x+ik_y),\,\,\,
\mathcal{H}^{BA}=\hbar v_F(k_x-ik_y),\,\,\,
\mathcal{H}^{AA}=t.
\end{equation}
The Fermi velocity, $v_F$, is defined by $\hbar v_F
= \sqrt{3}a\gamma_0/2$ and will be used as an intralayer parameter
instead of $\gamma_0$ in the subsequent consideration.

The effect of the magnetic field $\vec{B}=(0, B_y, B_z)$ is conveniently
introduced by Peierls substitution into the zero-field Hamiltonian. If
we choose the vector potential in the form
$\vec{A}=(B_yz-B_zy, 0, 0)$ we get
\begin{equation}
\hbar k_x \rightarrow \hbar k_x-|e|B_zy+|e|B_yjd.
\label{eqkx}
\end{equation}

When solving the Schr\"{o}dinger equation for the above four
wave functions, we can make use of the structure of $\mathcal{H}$, which
allows to express the function $\psi_j^B$ via the function $\psi_j^A$
from the same layer, and leaves us with two ,,interlayer'' equations
for $\psi_j^A$ and $\psi_{j+1}^A$.

Note, that graphite is no longer periodic in the $z$-direction
(parallel to the $c$-axis) but becomes periodic in the direction of
the tilted magnetic field. Here we apply an approach developed
in \cite{Goncharuk_2005}. The new periodicity implies that
$\psi_{j}^A$ can be written as
\begin{equation}
\psi_{j}^A=e^{i k_z d j}\phi_1^A\left(\eta+j\eta_d\right),\,\,\,\,
\psi_{j+1}^A=e^{i k_z d(j+1)}\phi_2^A\left(\eta+(j+1)\eta_d\right),
\label{eqperiot}
\end{equation}
$-\pi/2\leq k_z d\leq \pi/2$. Here $\phi_1^A\left(\eta+j\eta_d\right)$
and $\phi_2^A\left(\eta+j\eta_d\right)$ can be associated with cyclotron orbits in
two layers. The dimensionless variable $\eta$ is defined by $\eta =
(y-y_0)/\ell$, where $y_0$ is the cyclotron orbit center,
$y_0=\ell_z^2k_x$, and $\ell_z$ is the magnetic length, $\ell_z^2
= \hbar/(|e|B_z)$. The small parameter
\begin{equation}
\eta_d = \frac{B_y}{B_z}\frac{d}{\ell_z}
\end{equation}
means the shift of the cyclotron orbit center in the $j$-layer due to
the in-plane component of the magnetic field, $B_y$.
Introducing the shift operator by
\begin{equation}
\phi\left(\eta+\eta_d\right)=e^{i\kappa\eta_d}\phi(\eta),\,\,\,\
\kappa=-i\partial/\partial\eta,
\end{equation}
and employing $\kappa$-representation, two ,,interlayer'' equations
can be given the form
\begin{eqnarray}
\label{main-eq1}
\left[\frac{\mathcal{B}}{2}\left(-\frac{\partial^2}{\partial\kappa^2}
+\kappa^2+1\right)-E^2\right]
\phi_2^A+
2 t\, E\cos{(\kappa\eta_d+k_zd)} \phi_1^A=  0 ,\label{eq1tt} 
\\
\label{main-eq2}
\left[\frac{\mathcal{B}}{2}\left(-\frac{\partial^2}{\partial\kappa^2}
+\kappa^2
-1\right)-E^2\right]
\phi_1^A+
2 t\, E\cos{(\kappa\eta_d+k_zd)} \phi_2^A =  0,\label{eq2tt} 
\end{eqnarray}
where $\mathcal{B}$ stands for $2\hbar |e|v_F^2
B_z$. Eqs.~(\ref{main-eq1},\ref{main-eq2}) represent the main result
of this work.

It is evident that $\phi_1^A$ and $\phi_2^A$ cannot be too far from
the eigenfunctions of the harmonic oscillator, $\varphi_n(\kappa)$. 
If we write
\begin{equation}
\phi_1^A=\frac{1}{L_x}e^{ik_xx}\sum_{n'=0}^{\infty}A_{1,n'}\varphi_{n'},\,\,\,
\phi_2^A=\frac{1}{L_x}e^{ik_xx}\sum_{n'=0}^{\infty}A_{2,n'}\varphi_{n'},
\end{equation}
we arrive to
\begin{eqnarray}
\left[\mathcal{B}(n+1)-E^2\right] A_{2,n}+
2 E \sum_{n'=0}^{\infty}T_{n,n'}A_{1,n'}=0,\label{eq1p}
\\
\left[\mathcal{B}n-E^2\right] A_{1,n}+
2 E \sum_{n'=0}^{\infty}T_{n,n'}A_{2,n'}=0,\label{eq2p}
\end{eqnarray}
where
\begin{equation}
T_{n,n'}=
t\int_{-\infty}^{\infty}\varphi_{n}(\kappa)\cos(\kappa\eta_d+k_zd)
\varphi_{n'}(\kappa)
d\kappa.
\end{equation}

In the perpendicular magnetic field $\eta_d =0$ and
$T_{n,n'}=2\cos(k_zd)\delta_{n,n'}$. This corresponds to the
previously discussed case \cite{Koshino_2008}. At the $K$ point, 
$k_zd=0$, we get the energy spectrum of an effective bilayer. At the
$H$ point, $k_zd=\pm \pi/2$, the coupling between layers disappears, and we
obtain the LLs corresponding to graphene Dirac fermions,
namely $E_{1,n}^{\pm}=\pm\sqrt{\mathcal{B}n}$ for the first layer, and
$E_{2,n}^{\pm}=\pm\sqrt{\mathcal{B}(n+1)}$ for the second layer,
$n=0,1,2,\cdots$.

Let us discuss the magnetic field of an arbitrary direction. It is
obvious that the influence of the small parameter $\eta_d$ is
negligible at the point K when $k_zd=0$ and
$\cos{(\kappa\eta_d-k_zd)}\approx 1$. This case corresponds to the
bilayer subject to the tilted field discussed in Ref.~\cite{Hyun},
with the result that corrections induced by $B_y$ are negligible.

Let us consider the influence of $B_y$ at the H point, when
$\cos(k_zd)=0$ and $\sin(k_zd)=\pm 1$. Then, unlike the case of the
perpendicular field, the interlayer interaction is not reduced to
zero, but remains finite.
Note, that the small perturbation $\eta_d$ couples the unperturbed
states $|n_2^A\rangle$ and $|(n+1)_1^A\rangle$, which belong to the
same unperturbed eigenvalues $\pm\sqrt{\mathcal{B}(n+1)}$.
Consequently, the perturbation approach suitable to describe the coupling of
these degenerated states must be applied, which yields equations
\begin{eqnarray}
\label{eq1}
\left[\mathcal{B}(n+1)-E^2\right] A_{2,n}+2 E T_{n,n+1} A_{1,n+1}=0,\label{eq1pp}
\\
\label{eq2}
\left[\mathcal{B}(n+1)-E^2\right] A_{1,n+1}+2 E T_{n+1,n}A_{2,n}=0,\label{eq2pp}
\end{eqnarray}
where
\begin{equation}
T_{n,n+1}=T_{n+1,n}= t \left[\eta_d \sqrt{\frac{n+1}{2}}-\frac{1}{2}
\left(\eta_d \sqrt{\frac{n+1}{2}}\right)^3 + \ldots \right].
\end{equation}

The secular equation derived from Eqs.(\ref{eq1},\ref{eq2}) reads
\begin{equation}
\left[\mathcal{B}(n+1)-E^2\right] ^2 - 4 E^2 T^2_{n,n+1}=0,
\end{equation}
and from here we get the four eigenenergies
\begin{equation}
E_{n+1}^{\pm,\pm}=\pm T_{n,n+1}\pm\sqrt{\mathcal{B}(n+1)+T^2_{n,n+1}},
\,\,\, n = 0,1,2,\cdots.
\end{equation}
The eigenenergies,  $E_{0}^{\mp}$ and $E_{0}^{\pm}$, originated from 
$E_{1,0}^{\pm}$ remain the same as in the perpendicular magnetic
field. In that case the degeneracy is not removed.
\begin{figure}
\label{fig}
 \includegraphics[width=\textwidth]{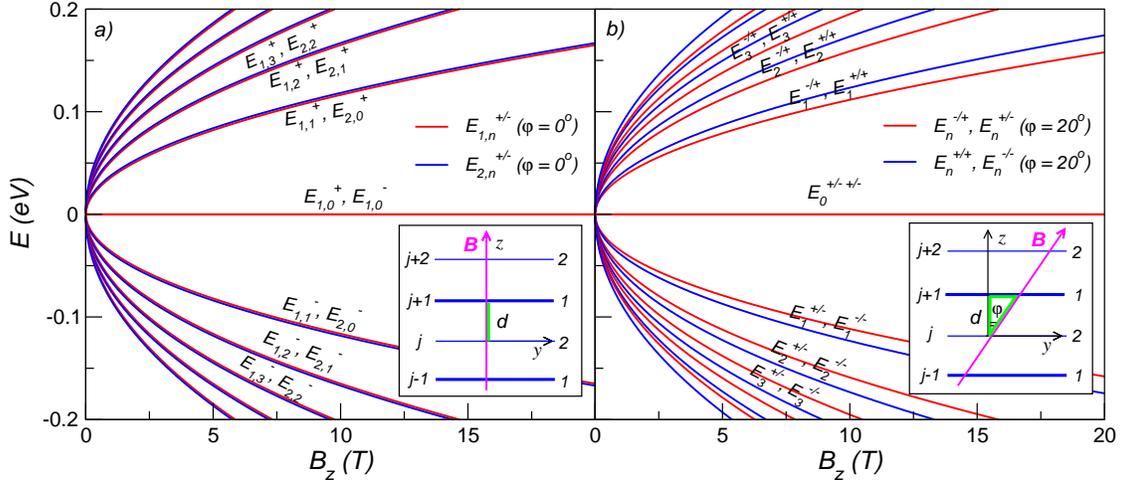}
 \caption{LLs of graphite   at the H point as a function of the
 perpendicular component of the magnetic field, $B_z$. 
{\emph a)} LLs,  $E_{1,n}^{\pm}$ and $E_{2,n}^{\pm}$, in the perpendicular magnetic field ($\varphi  = 0^o$). {\emph b)} LLs,  $E_{n}^{\pm,\pm}$, splitted by  tilting the magnetic field ($\varphi = 20^o$).}
\end{figure}
Lifting of LL degeneracy by the tilted magnetic
field is shown in Fig.~\ref{fig}. The LL splitting is of the
order of several meV, and it grows with the tilt angle, i.e., with $B_y$. 

To conclude, we have obtained the LL structure of graphite at the H
point of the Brillouine zone in the tilted magnetic field
configuration. This effect is experimentally observable.


\begin{theacknowledgments}
The authors benefited from discussions with Milan Orlita.  The support
of the European Science Foundation EPIGRAT project (GRA/10/E006), AV
CR research program AVOZ10100521 and the Ministry of Education of the
Czech Republic project LC510 is acknowledged.
\end{theacknowledgments}
\bibliographystyle{aipproc}   
\bibliographystyle{aipprocl} 


\end{document}